\RequirePackage{lineno}
\documentclass[aps,titlepage,11pt,superscriptaddress]{revtex4}
\usepackage{epsfig}
\usepackage{float}
\usepackage{graphicx}
\usepackage{times}
\usepackage{amsmath}
\usepackage{color}
\usepackage{lineno}

\begin{document}

%\linenumbers

\title{Evolution of conditional cooperation in collective-risk social dilemma with repeated group interactions}

\author{Shijia Hua}
\affiliation{College of Science, Northwest A \& F University, Yangling 712100, China}

\author{Zitong Hui}
\affiliation{College of Science, Northwest A \& F University, Yangling 712100, China}

\author{Linjie Liu}
\email{linjieliu1992@nwafu.edu.cn}
\affiliation{College of Science, Northwest A \& F University, Yangling 712100, China}

\begin{abstract}\noindent
\\
The evolution and long-term sustenance of cooperation has consistently piqued scholarly interest across the disciplines of evolutionary biology and social sciences. Previous theoretical and experimental studies on collective risk social dilemma games have revealed that the risk of collective failure will affect the evolution of cooperation. In the real world individuals usually adjust their decisions based on environmental factors such as risk intensity and cooperation level. However, it is still not well understood how such conditional behaviors affect the evolution of cooperation in repeated group interactions scenario from a theoretical perspective. Here, we construct an evolutionary game model with repeated interactions, in which defectors decide whether to cooperate in subsequent rounds of the game based on whether the risk exceeds their tolerance threshold and whether the number of cooperators exceeds the collective goal in the early rounds of the game. We find that the introduction of conditional cooperation strategy can effectively promote the emergence of cooperation, especially when the risk is low. In addition, the risk threshold significantly affects the evolutionary outcomes, with a high risk promoting the emergence of cooperation. Importantly, when the risk of failure to reach collective goals exceeds a certain threshold, the timely transition from a defective strategy to a cooperative strategy by conditional cooperators is beneficial for maintaining high-level cooperation.
\end{abstract}

\maketitle

\noindent

Cooperation is a fundamental aspect of both natural systems and human societies \cite{Perc_PR_17,Panis_22,Stockley_11,Thomas_04}. In the natural world, many species engage in cooperative behaviors such as hunting \cite{Boesch_94}, foraging \cite{Ranta_93}, and caring for offspring \cite{Wilkinson_16}.  Similarly, in human societies, cooperation is essential for achieving common goals such as preventing the spread of infectious diseases \cite{chen_22prsa,chen_19prsb}, mitigating climate change \cite{Andrews_18,Grimalda_2022}, and maintaining public resources \cite{Schlter_16}. However, maintaining effective public cooperation among genetically unrelated individuals can be challenging due to various factors. The failure of cooperation can often be attributed to the inherent conflict between the maximization of self-interest and the maximization of group interest. Individuals tend to prioritize their own interests, often at the expense of the group's overall interests \cite{Bornstein_94}.

Evolutionary game theory offers a theoretical scaffold to probe the rise of cooperation, especially when kin selection is non-functional, as observed among genetically unrelated individuals \citep{han2015,han2021,han2022,han2022interface,Liu_22interface,Tanimoto2017,Tanimoto2015,Tanimoto2021,Szolnoki2014interface}. 
In recent years, various theoretical models such as the prisoner's dilemma game, snowdrift game, stag hunt game, and public goods game have been proposed to study the evolution of cooperation in real-world scenarios \cite{Xia2018NJP,Zhu2022,xia2022}. In addition to these models, the collective-risk social dilemma game, as a nonlinear public goods game, is gaining recognition due to its potential for application in climate change and migration \cite{Chen2012,He2019,chen2014,chen2012epl,wang2009PRE,Milinski2008,Santos2012,Szekely2021,Abou2018}. In this model framework, all participants in the game have an initial endowment, and cooperators contribute a fraction of their initial endowment, while defectors do not contribute. If the number of cooperators in the game group does not reach the collective goal, all individuals will lose all their remaining endowments with a probability of $r$, otherwise each individual will retain their own endowment \cite{Santos2012,santos_11}. Previous experimental and theoretical studies have revealed that high risk stimulates individuals' willingness to cooperate, thereby promoting the achievement of collective goals \cite{Milinski2008,Liu2018,sun2021iScience,gis2019sr,Vasconcelos2014pnas}. Recent studies have incorporated factors such as costly punishment \cite{Jiang2023chaos} and communication \cite{Wang2020pnas} into the collective-risk social dilemma game model and also applied it to address carbon emissions.

Although a considerable amount of research has been conducted on the collective-risk social dilemma game, there are still many questions that warrant further exploration. One key issue to consider is how conditional cooperation strategies influence the evolution of cooperation in repeated group interactions. Previous studies have primarily focused on one-shot game scenarios \cite{Chen2012,He2019,chen2014,chen2012epl,wang2009PRE}, whereas in reality, interactions between individuals often occur repeatedly \cite{Van Segbroeck_12PRL,Liu_22interface,Liu_prsa_22}. In such repeated interactions, where the game is played over many rounds, individuals are able to adjust their behavioral decisions based on the game environment they find themselves in.  Hilbe and coworkers \cite{Hilbe13plos} employed evolutionary game theory to investigate the impact of strategic timing on the evolutionary outcomes in collective-risk social dilemmas. In their model, individuals were given the choice to wait for others' decisions or influence others by taking the lead. The findings suggest that the timing of participants' contributions significantly increased the probability of moving towards efficient equilibrium. Furthermore, in Abou Chakra and Traulsen's study \cite{Chakra12pcb}, each participant determined their contribution ($a$ or $b$) based on whether the total sum exceeded their threshold. Under the all of nothing piecewise risk function they saw the players that delayed their contributions were favored in high risk setting.

In addition to the theoretical studies mentioned above, conditional behavior in repeated interaction scenarios is also very common in real-life situations. Taking the example of climate summits, countries also decide whether to take cooperative action based on their own interests and risk preferences \cite{Tingley_14}. If a country perceives significant risks to its national security and economic development from climate change and believes that taking action can mitigate these risks, it may be inclined to take cooperative action, such as signing a climate agreement and committing to reducing greenhouse gas emissions. However, if a country believes that taking action would have negative impacts on its economic interests or does not trust that other countries will also take action, it may choose not to cooperate. Therefore, a country's decision to cooperate is often based on its perception and expectation of risks and cooperative actions. Along these lines, prior studies have incorporated wealth inequalities \cite{Abou2018,Vasconcelos2014pnas,JTheoreticalBiology_2014} and heterogeneity in risks \cite{Abou2018} into the collective risk social dilemma, revealing that the importance of capturing real-life scenarios for individual decision-making. For example, Abou Chakra and Traulsen \cite{JTheoreticalBiology_2014} found that the poor contribute only when early contributions are made by the rich players. Abou Chakra and coworkers \cite{Abou2018} found that `wait and see' strategy is effective only when players are aware of the critical time to contribute to avert danger, and if their contributions can effectively mitigate the risks. 

In this work, our aim is to construct a game-theoretic model based on the collective-risk social dilemma. We investigate the evolutionary dynamics of conditional cooperation strategy in a scenario of repeated group interactions. In this scenario, defectors observe the game environment to assess the level of risk and cooperation within the group during the early rounds of the game. Based on this assessment, they make decisions in subsequent rounds on whether to cooperate. These decisions are influenced by factors such as whether the risk exceeds their tolerance threshold and whether the group has achieved the desired level of cooperation in relation to the collective goal.

\vbox{}

\leftline{\textbf{Model and Methods}}

We consider a well-mixed population from which $N$ individuals are randomly selected to participate in a collective risk social dilemma game \cite{Santos2012}. Each player has an initial endowment of $b$ and can decide whether to contribute towards achieving collective goals. Cooperators ($C$) incur a cost of $c$, while defectors ($D$) contribute nothing. If the number of cooperators in the game group is less than the collective goal $n_{pg}$, all individuals will lose their endowments with probability $r$. Otherwise, all individuals maintain their endowments. We further introduce repeated group interactions, where the game is repeated with probability $w$, which leads to an expected number of game rounds $F = 1/(1-w)$ \cite{Liu_22interface,Sigmund10CS}.

Under the framework of repeated collective-risk social dilemma game, we introduce conditional cooperation strategy, where all individuals engage in collective-risk social dilemma game in the first $\xi-1$ rounds. During these periods, defectors incur a cost of $\sigma$ to observe the game environment, specifically, they can know about the risk level and cooperation level of the group in the first $\xi-1$ rounds. Defectors then evaluate the situation by comparing the risk level of the game with their own tolerance threshold $T$, and the number of cooperators in the group with the collective target $n_{pg}$, and decide whether to cooperate. Only when both conditions are met, will defectors choose to cooperate in subsequent rounds of the game. In order to provide a clearer understanding of our model configuration, we present our model parameters and their meanings in Table 1.

\begin{table}[H]
\centering
\caption{Model parameters and their corresponding definitions}
\begin{tabular}{l|c}
Parameter & Meaning\\\hline
$b$ & Initial endowment \\
$c$ & Cost of cooperation \\
$r$ & Risk \\
$n_{pg}$ & Collective goal \\
$N$ & Group size \\
$N_{C}$ &Number of cooperators in the group \\
$T$ &  Tolerance threshold\\
$w$ & Repeated game probability  \\
$\xi$ & Rounds for conditional cooperators to switch strategies \\
$F$ & Expected total number of game rounds \\
$\sigma$ & Observation cost
\\\hline
\end{tabular}\\
\end{table}\label{table1}

Firstly, we consider a population composed of cooperators and conditional cooperators. According to the above description, we can write the payoffs for cooperators and conditional cooperators as follows:
\begin{eqnarray*}
\pi_{C}&=&
\begin{cases}
    [-c+(1-r)b](\xi-1)+(b-c)(F-\xi+1),\quad \text{if } N_{C}+1<n_{pg} \quad \&  \quad r\geq T \\
    \{-c+b\theta(N_{C}+1-n_{pg})+(1-r)b[1-\theta(N_{C}+1-n_{pg})]\}F, \quad \text{otherwise}
\end{cases}\\
\pi_{DC}&=&
\begin{cases}
    (1-r)b(\xi-1)+(b-c)(F-\xi+1)-\sigma, \quad \text{if } N_{C}<n_{pg} \quad \&  \quad r\geq T \\
    \{b\theta(N_{C}-n_{pg})+(1-r)b[1-\theta(N_{C}-n_{pg})]\}F-\sigma, \quad \text{otherwise,}
\end{cases}
\end{eqnarray*}
where the terms $[-c+(1-r)b](\xi-1)$ and $(1-r)b(\xi-1)$ denote the payoffs of cooperators and conditional cooperators (defectors) in the first $\xi-1$ rounds of the game when the number of cooperators in the group is below the collective target and the risk is not below the tolerance threshold. $(b-c)(F-\xi+1)$ represents the payoff of cooperators and conditional cooperators in the subsequent $F-\xi+1$ rounds of the game when the number of cooperators in the group is below the collective target and the risk is not below the tolerance threshold. Otherwise, the payoffs of cooperators and conditional cooperators (defectors) are the payoffs of $F$ rounds repeated collective-risk social dilemma game.

Next, we consider a population consisting of cooperators, defectors, and conditional cooperators. Accordingly, we can express the payoffs of cooperators, defectors, and conditional cooperators as follows:
\begin{eqnarray*}
\pi_{C}&=&
\begin{cases}
    [-c+(1-r)b](\xi-1)+[b\theta(N_{C}+N_{DC}+1-n_{pg})-c+\\(1-r)b(1-\theta(N_{C}+N_{DC}+1-n_{pg}))](F-\xi+1),\quad \text{if } N_{C}+1<n_{pg} \quad \&  \quad r\geq T \\
    \{-c+b\theta(N_{C}+1-n_{pg})+(1-r)b[1-\theta(N_{C}+1-n_{pg})]\}F, \quad \text{otherwise}
\end{cases}\\
\pi_{D}&=&
\begin{cases}
    (1-r)b(\xi-1)+[b\theta(N_{C}+N_{DC}-n_{pg})+\\
(1-r)b(1-\theta(N_{C}+N_{DC}-n_{pg}))](F-\xi+1), \quad \text{if } N_{C}<n_{pg} \quad \&  \quad r\geq T \\
    \{b\theta(N_{C}-n_{pg})+(1-r)b[1-\theta(N_{C}-n_{pg})]\}F, \quad \text{otherwise}
\end{cases}\\
\pi_{DC}&=&
\begin{cases}
    (1-r)b(\xi-1)+[b\theta(N_{C}+N_{DC}+1-n_{pg})-c+\\
(1-r)b(1-\theta(N_{C}+N_{DC}+1-n_{pg}))](F-\xi+1)-\sigma, \quad \text{if } N_{C}<n_{pg} \quad \&  \quad r\geq T \\
    \{b\theta(N_{C}-n_{pg})+(1-r)b[1-\theta(N_{C}-n_{pg})]\}F-\sigma, \quad \text{otherwise,}
\end{cases}
\end{eqnarray*}
where $N_{DC}$ denotes the number of conditional cooperators in the group.

In the following, we investigate the evolutionary dynamics of the system in an infinite well-mixed populations by analyzing replicator equations \cite{Sigmund10CS,Schuster_83,Hofbauer_98}. Subsequently, we employ the Markov decision process to address the stochastic effects and population dynamics in a finite well-mixed population \cite{Imhof_05,Kampen_07}.

\leftline{\textbf{Replicator equation}}
In an infinite population, we consider a population state as $\mathbf{x}=(x_1, x_2, \dots, x_n)$ where $x_i$ denotes the frequency of adoption of $i$ strategy by individuals in the population. The rate of change of the frequency of a strategy in a population is proportional to the difference between the average payoff of that strategy and the average payoff of the population as a whole. Accordingly, the replicator equation can be written as
\begin{eqnarray}
\dot{x}_{i}=x_{i}(\Pi_{i}(\mathbf{x})-\bar{\Pi}(\mathbf{x})),
\end{eqnarray}
where $\Pi_{i}(\mathbf{x})$ denotes the average payoff of an individual using strategy $i$ and $\bar{\Pi}(\mathbf{x})=\sum_{j=1}^{n}x_{j}\Pi_{j}(\mathbf{x})$ denotes the average payoff of the whole population. The direction of change for each strategy can be represented by the differential equations above, which allows for a complete characterization of the strategic variations in the population.

\vbox{}

\leftline{\textbf{Markov decision process}}

When individuals interact in finite populations, stochastic effects including behavioral mutations and imitation errors become non-negligible. The stochastic dynamics in such finite populations can be described by the gradient of selection, defined as the difference between the probabilities of increasing and decreasing the number of given strategies, and the stationary distribution of the associated Markov chain, which characterizes the pervasiveness in time of a given composition of the population \cite{Vasconcelos_13}.

We adopt a pairwise comparison rule to describe the process of strategy selection. Specifically, at each time step, an individual $A$ is randomly selected to update its strategy, and with probability $\frac{1}{1+\exp[\beta(f_{A}-f_{B})]}$, $A$ imitates the strategy of another randomly selected individual $B$, where $\beta$ is referred to as the intensity of selection \cite{szab_98}, and $f_{A}$ and $f_{B}$ represent the average payoffs of individuals $A$ and $B$, respectively. Furthermore, we introduce behavioral mutation, where with probability $\mu$, individual $A$ randomly selects a strategy from the remaining strategy space to use, and with probability $1-\mu$, $A$ updates its strategy using the pairwise comparison rule described above. Concretely, in a population with $n$ strategies, the probability that individual $A$ adopts the strategy of individual $B$ can be written as
\begin{eqnarray}\label{trans}
T_{A\rightarrow B}&=&(1-\mu)\frac{i_{A}}{Z}\frac{i_{B}}{Z-1}\frac{1}{1+\exp[\beta(f_{A}-f_{B})]}+\mu\frac{i_{A}}{(n-1)Z},
\end{eqnarray}
where $i_{A}$ and $i_{B}$ denote the numbers of $A$ and $B$ individuals in finite populations, respectively. 

Therefore, in a tri-strategy population, the gradient of selection \cite{Vasconcelos_13}, used to characterize the most probable direction of system evolution after leaving the current configuration $\textbf{i}=(i_{A}, i_{B})$ can be expressed as:
$\nabla_{\textbf{i}}=(T_{\textbf{i}}^{A+}-T_{\textbf{i}}^{A-})\textbf{u}_{\textbf{A}}+(T_{\textbf{i}}^{B+}-T_{\textbf{i}}^{B-})\textbf{u}_{\textbf{B}},$
where $\textbf{u}_{\textbf{A}}$ and $\textbf{u}_{\textbf{B}}$ are basis vectors. $T_{\textbf{i}}^{A+} (T_{\textbf{i}}^{A-})$ and $T_{\textbf{i}}^{B+} (T_{\textbf{i}}^{B-})$ respectively denote the probabilities that the numbers of $A$ and $B$ individuals increase (decrease) one. In a bi-strategy population, the gradient of selection is defined as the difference between the probabilities of increasing and decreasing the number of individuals adopting a given strategy, namely, $G(i_{A})=T^{+}(i_{A})-T^{-}(i_{A})$.

Based on Eq. (\ref{trans}), we can derive the transition probability from state $\textbf{i}$ to an adjacent state $\textbf{i}^{'}$. Thus, the transition matrix $T=[T_{\textbf{i},\textbf{i}^{'}}]$, which fully characterizes the stochastic process dynamics.
The stationary distribution of the system, representing the amount of time the population spends at each state, can be analytically computed by normalizing the eigenvector associated with the eigenvalue 1 of the transition matrix $T$ of the Markov Chain \cite{Vasconcelos_13}.

\vbox{}

\leftline{\textbf{Results}}
\leftline{\textbf{Evolutionary dynamics in infinite populations}}

We first consider a population consisting only of cooperators and conditional cooperators, where the fraction of cooperators is $x$ and the fraction of conditional cooperators is $1-x$. In an infinite well-mixed population, we use the gradient of selection to characterize the evolution of strategies. Concretely, the replicator equation can be expressed as $\dot{x}=x(1-x)(\Pi_{C}-\Pi_{DC})$, where $\Pi_{C}$ and $\Pi_{DC}$ respectively denote the expected payoffs of cooperators and conditional cooperators, and $\dot{x}$ means the rate of change in $x$ over time.

\begin{figure}[t]
\centering
\includegraphics[width=1\linewidth]{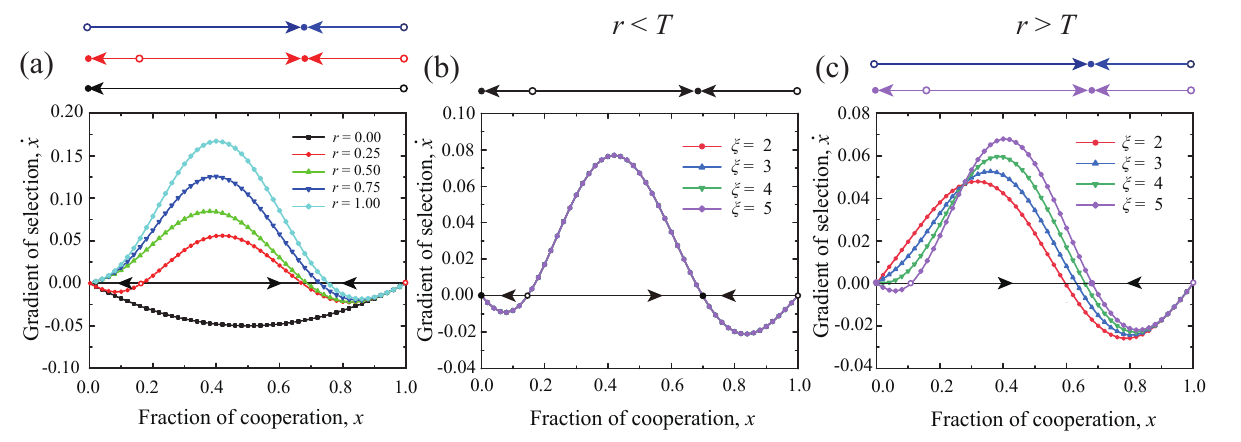}
\caption{Gradient of selection of the replicator dynamics. Panel (a) shows that high risk can promote high-level cooperation. Panel (b) shows that the number of rounds in which the conditional cooperator adjusts its strategy does not affect the outcome when the risk is below the tolerance threshold. Panel (c) reveals that timely adjustment of the strategy is advantageous for maintaining cooperation when risk exceeds the tolerance threshold. Parameters are $N=6, n_{pg}=3, \xi=3, T=0.5, w=0.8$, and $\sigma=0.3$ in panel (a); $N=6, n_{pg}=3, r=0.3, T=0.5, w=0.8$, and $\sigma=0.3$ in panel (b); $N=6, n_{pg}=3, r=0.3, T=0.2, w=0.8$, and $\sigma=0.3$ in panel (c).}
\label{fig1}
\end{figure}

According to the replicator equation, when $\dot{x} > 0$, the frequency of cooperators in the population will increase. In Fig. 1(a), we present the result that the gradient of selection changes with the fraction of cooperators in the population for different risk values. In the absence of risk, the gradient value $\dot{x}$ is always negative, resulting in conditional defectors occupying the entire population regardless of the initial conditions. However, with a slight increase in the risk value $(r=0.25)$, we observe the emergence of two internal equilibrium points, one of which is stable and the other is unstable. This risk value was previously considered unable to promote the emergence of cooperation in theoretical study \cite{santos_11}. When the risk value is high $(r \geq 0.5)$, the unstable equilibrium point disappears, and sustained cooperation at a significant level can be maintained. This means that an increase in risk is conducive to the emergence of a high percentage of cooperation $(x>0.6)$. Furthermore, we investigate the impact of the game round $\xi$ that conditional cooperators adjust their strategy on the evolutionary outcomes when the risk value is below the perception threshold $(r<T)$ and above the perception threshold $(r>T)$, respectively. We find that when the risk value is below the tolerance threshold, different values of $\xi$ do not affect the evolutionary outcome, where the system exhibits an unstable equilibrium point and a stable equilibrium point (see Fig. 1(b)). When the risk exceeds the tolerance threshold, we find that an intermediate value of $\xi$ can sustain a high level of cooperation. However, when $\xi$ is particularly high ($\xi=5$), the system exhibits bistability, where depending on the initial conditions, the system either converges to a high-level cooperation state or the $D$ state (see Fig. 1(c)). This means that an individual choosing a free-riding strategy can also be favored by natural selection.

\begin{figure}[t]
\centering
\includegraphics[width=1\linewidth]{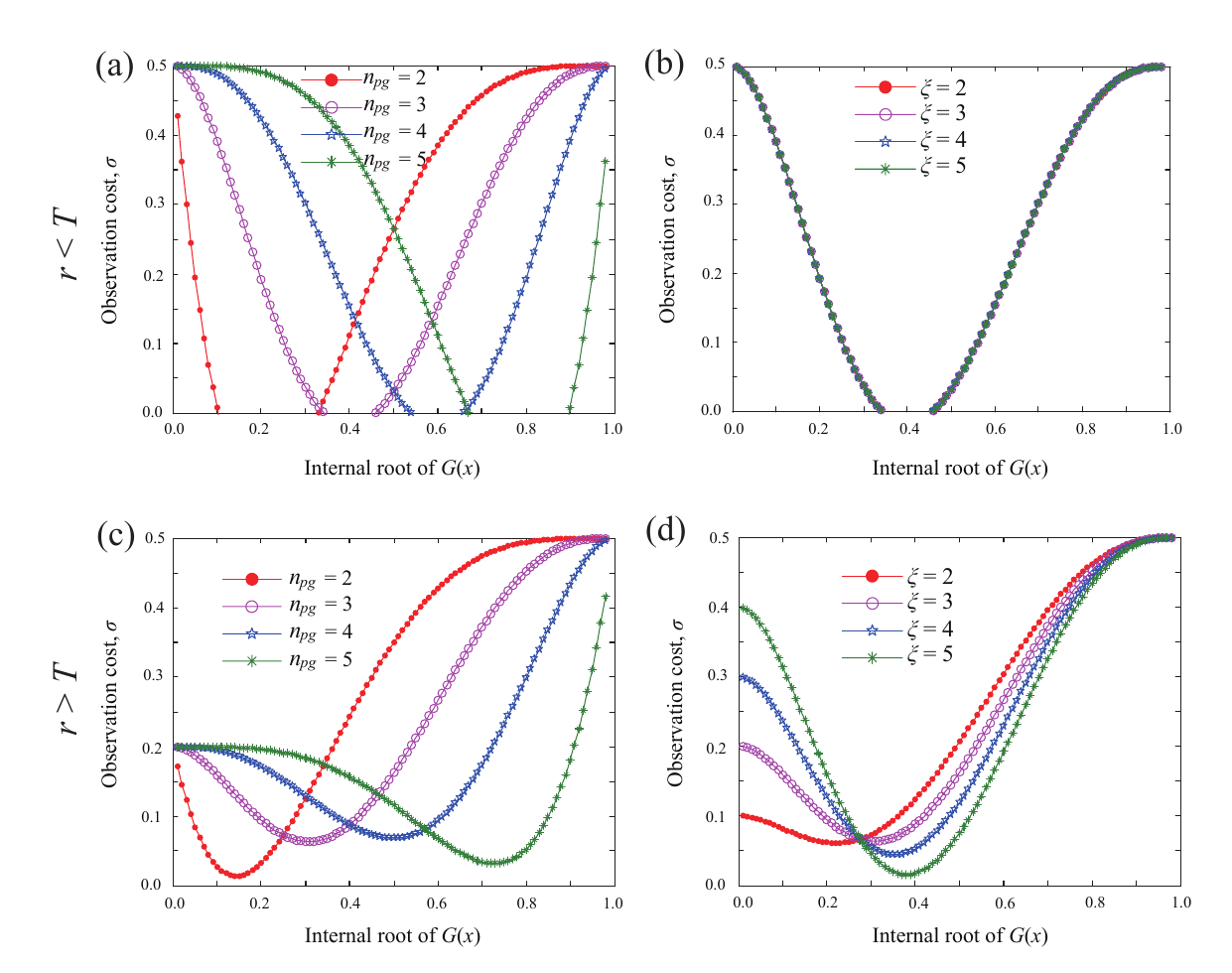}
\caption{Internal roots of the gradient of selection as a function of observation cost for different $n_{pg}$ and $\xi$ values. Expanding the observation cost will enhance the evolutionary advantage of cooperation. Parameters are $N=6, \xi=3, T=0.5, w=0.8, r=0.3$ in panel (a); $N=6, n_{pg}=3, r=0.3, T=0.5, w=0.8$ in panel (b); $N=6, r=0.3, T=0.2, w=0.8, \xi=3$ in panel (c); $N=6, r=0.3, T=0.2, w=0.8, n_{pg}=3$ in panel (d).}
\label{fig2}
\end{figure}

In Fig. 2, we show the location of internal equilibria as a function of observation cost $\sigma$ for different values of $\xi$ and $n_{pg}$ when the risk value is lower than the tolerance threshold and is higher than the tolerance threshold, respectively. We observe that as the value of $\sigma$ increases, the value of the unstable internal equilibrium point gradually decreases until it reaches zero, while the stable equilibrium point gradually increases until it reaches one. This indicates that the increase in observation cost expands the basin of attraction of stable equilibrium point, thereby enabling the maintenance of high-level cooperation. Besides, the increase in collective goals leads to an increase in the values of existing internal equilibrium points, regardless of whether the collective risk exceeds the tolerance threshold of conditional cooperators (see Fig. 2(a) and (c)). In addition, the increase of collective goals reduces the attraction domain of stable internal equilibrium point, which is therefore detrimental to the maintenance of high-level cooperation. Furthermore, we find that when the risk is below the tolerance threshold, the increase in the value of $\xi$ has no effect on the internal equilibrium points (see Fig. 2(b)). However, when the risk exceeds the tolerance threshold, we observe that an increase in the value of $\xi$ leads to an increase in the values of internal equilibrium points (see Fig. 2(d)), which implies that the attraction domain of stable equilibrium points decreases, thereby detrimental to the maintenance of cooperation.

\begin{figure}[t]
\centering
\includegraphics[width=1\linewidth]{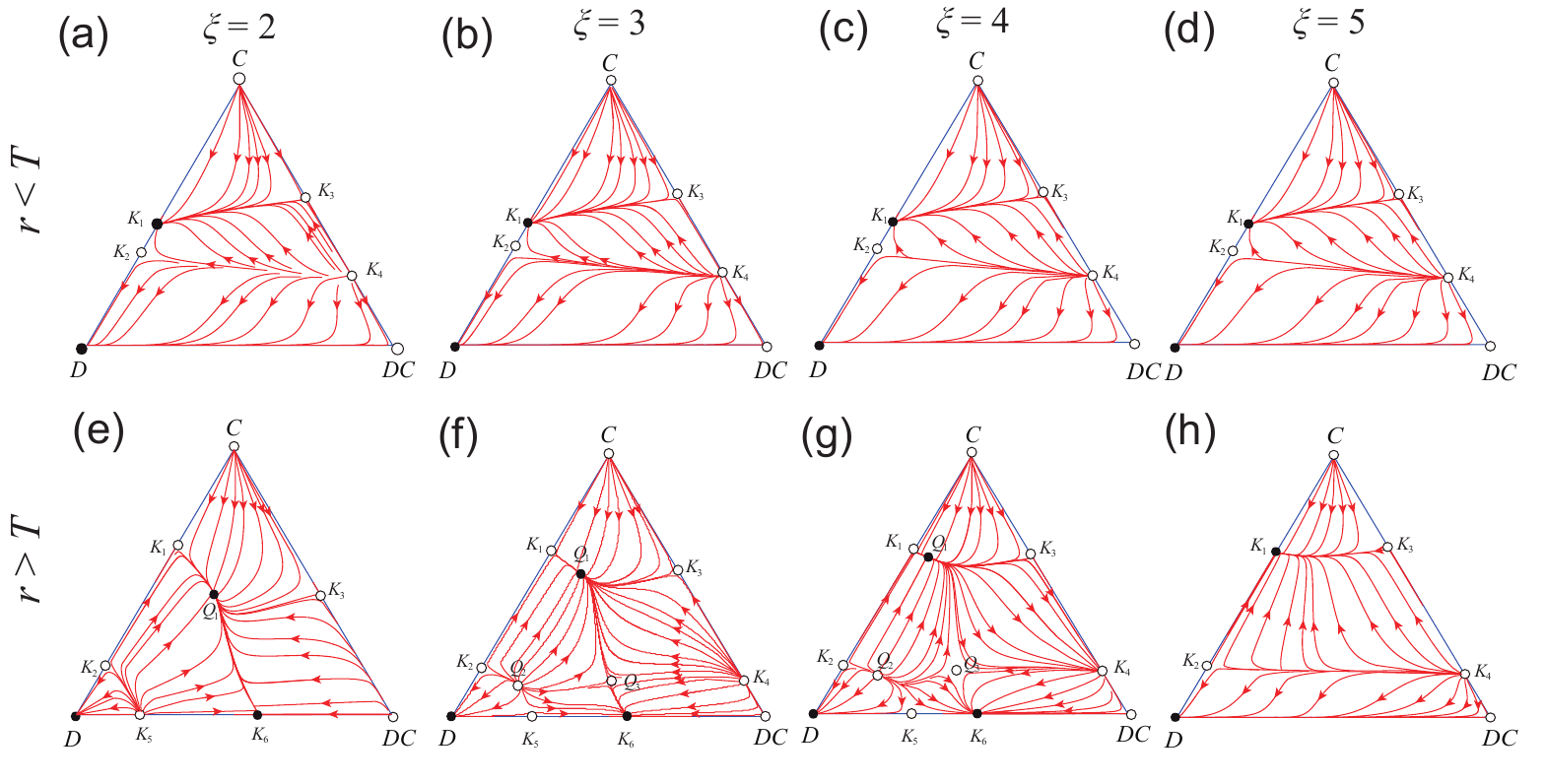}
\caption{Evolutionary dynamics of cooperation, defection, and conditional cooperation strategies in the simplex $S_3$. Empty circles represent unstable equilibrium points, while solid dots represent stable equilibrium points. Arrows represent the direction of evolution. Cooperators are able to stably coexist with defectors and conditional cooperators in a population when the risk exceeds the threshold of tolerance. Parameters are $N=6, n_{pg}=3, c=0.1, b=1, T=0.5, r=0.3, w=0.8, \sigma=0.1,$ and $\xi=2$ in panel (a); $\xi=3$ in panel (b); $\xi=4$ in panel (c); $\xi=5$ in panel (d); $N=6, n_{pg}=3, c=0.1, b=1, T=0.2, r=0.5, w=0.8, \sigma=0.1,$ and $\xi=2$ in panel (e); $\xi=3$ in panel (f); $\xi=4$ in panel (g); $\xi=5$ in panel (h).}
\label{fig3}
\end{figure}

Next, we consider a population consisting of cooperators, defectors, and conditional cooperators, where the frequency of cooperators is $x$, the frequency of defectors is $y$, and the frequency of conditional cooperators is $z$. In Fig. 3, we numerically investigate the impact of different values of $\xi$ on the evolutionary dynamics of cooperators, defectors, and conditional cooperators. The top row shows evolutionary results of the system when the risk is lower than the tolerance threshold. There exist seven equilibrium points, among which the vertex equilibrium point $(x,y,z)=(0,1,0)$ and the $D-C$ boundary equilibrium point $K_{1}$ are stable. Therefore, when the initial fraction of cooperation is not particularly low, cooperators can stably coexist with defectors in the population. Moreover, as the value of $\xi$ increases, the evolutionary outcomes of the system remain unchanged (see Fig 3(a)-(d)). The reason for this phenomenon is that the risk has not reached the tolerance threshold of conditional cooperators, and therefore they will not change their own strategies.

When the risk exceeds the tolerance threshold, the system could generate new evolutionary dynamics. Specifically, when the value of $\xi$ is small ($\xi=2$), we find that there are nine equilibrium points in the system, among which the interior equilibrium point $Q_{1}$, vertex $D$, and equilibrium point $K_{6}$ on the $D-DC$ boundary are stable (see Fig. 3(e)). The majority of interior trajectories converge to the interior stable point, thus cooperators, defectors, and conditional cooperators can stably coexist in the population. As the value of $\xi$ increases, we observe that the interior equilibrium point moves closer to the boundary $D-C$, and two new interior equilibrium points, $Q_{2}$ and $Q_{3}$, can emerge, which results in a reduction of the basin of attraction of the interior equilibrium point $Q_{1}$ (see Fig 3(f) and (g)). When the value of $\xi$ is extremely large ($\xi=5$), we observe the disappearance of the interior equilibrium point, resulting in the existence of seven equilibrium points in the system (see Fig 3(h)). Among them, stable equilibrium is achieved at point $K_{1}$ on the boundary of $D-C$ and at vertex $D$. It should be noted that the basin of attraction of $K_{1}$ exceeds that of vertex $D$, indicating that high-level cooperation can still be sustained, attributed to the high risk level. It should be noted that the attraction domain of $D$ is also the largest compared to the previous situations.

We further investigate the impact of other model parameters on the evolutionary outcomes. In Fig. A1, we show the role of different levels of risk in the evolution of cooperation, and find that cooperation cannot emerge in scenario with low risk, while increasing risk can effectively promote the emergence of cooperation. Specifically, when the risk level exceeds the tolerance threshold, there exists a stable interior equilibrium point, where the three types of strategists can stably coexist in the population. Moreover, we find that increasing the collective target value ($n_{pg}$) expands the attraction domain of full defection (see Fig. A2). It is worth emphasizing that when the risk exceeds the tolerance threshold, the system still exhibits a stable interior equilibrium point, although its attraction domain decreases with increasing values of $n_{pg}$, it is still larger than the attraction domain of $D$, thus enabling the maintenance of high-level cooperation. We are further interested in the impact of observation cost $\sigma$ on the evolutionary outcomes. As shown in Fig. A3, we present the effects of different observation costs on the evolution of cooperation in scenarios where the risk is lower than the tolerance threshold ($r<T$) and higher than the tolerance threshold ($r>T$), respectively. We find that when $r<T$, the system exhibits bistability, where the trajectory of the system converges to either the vertex $D$ or the point $K_{1}$ on the $C-DC$ boundary, depending on the initial conditions. Besides, increasing the observation cost $\sigma$ alters the distribution of equilibrium points on the $C-DC$ boundary. When $r > T$, the system can exhibit a stable interior equilibrium point when observation cost is low. As observation cost increases, the interior equilibrium point disappears and most system trajectories converge to the equilibrium point $K_{1}$ on the $D-C$ boundary. This implies that cooperators and defectors can stably coexist in the population. We also further investigated the impact of the repeated probability of game on the evolutionary results. As shown in Fig. A4, when the risk is below the tolerance threshold, we find that depending on the initial proportion of each strategy, the system will converge to a state of coexistence between cooperators and defectors or a state where defectors dominate. The repeated probability of game does not have a significant impact on the evolutionary outcome (see Fig. A4 (a)-(d)). When the risk is above the tolerance threshold, cooperators, defectors, and conditional cooperators can stably coexist in the population. The increase of repeated probability promotes the emergence of this stable coexistence state (see Fig. A4 (e)-(h)).

\vbox{}
\leftline{\textbf{Evolutionary dynamics in finite well-mixed populations}}

\begin{figure}[t]
\centering
\includegraphics[width=1\linewidth]{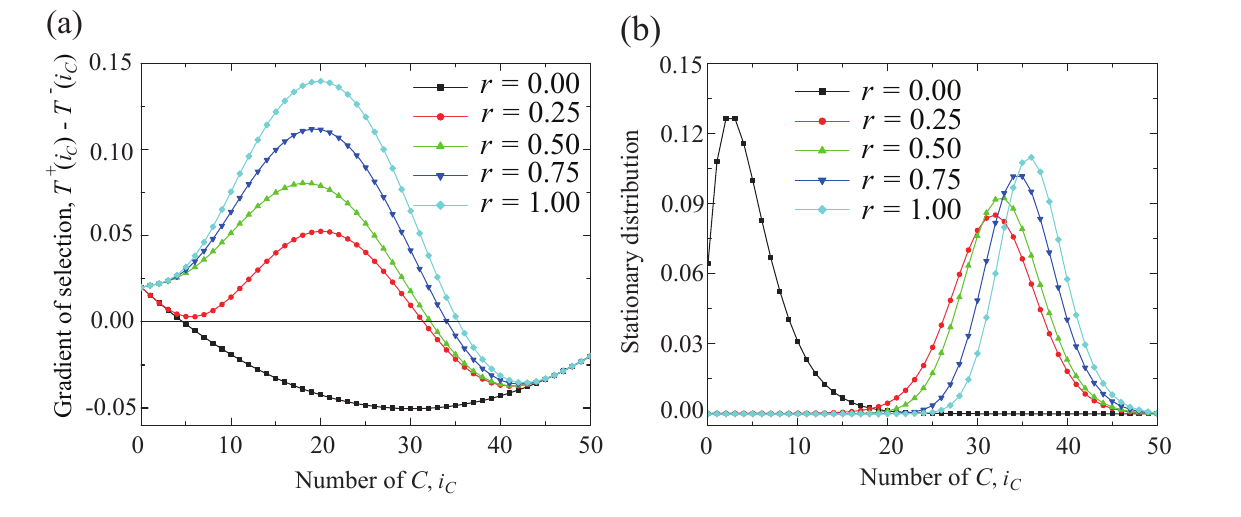}
\caption{Evolutionary dynamics of cooperation and conditional cooperation in finite populations. Panel (a) presents the gradient of selection as a function of the number of cooperators for different risk. Panel (b) displays the stationary distribution for a finite population to describe the prevalence of each state. High risk promotes the emergence of cooperation. Parameters are $Z=50, N=6, n_{pg}=3, \xi=3, T=0.5, w=0.8, \mu=1/Z, \beta=2$, and $\sigma=0.3$.}
\label{fig4}
\end{figure}

When the population size is finite, we employ the Markov process to analyze the evolutionary dynamics of the system. In Fig. 4, we present the results of the gradient of selection and stationary distribution when the population is composed of cooperators and conditional cooperators. Due to the existence of behavioral mutation, we find that in the absence of risk, the gradient equation $G(i_{C})=0$ has a stable internal root, in which cooperators and conditional cooperators can coexist stably within the population (see Fig. 4(a)). As the level of risk increases, we observe that the proportion of cooperators at the stable equilibrium further increases. Importantly, higher levels of risk ensure a higher fraction of cooperators at the stable equilibrium. In Fig. 4(b), we show the stationary distributions for different values of risk. In the absence of risk, the population spends most of its time in the state of defection. However, as the level of risk increases, the population spends a significant amount of time in the configuration where cooperation prevails, regardless of initial conditions. Similarly, higher levels of risk ensure a high level of cooperation.

\begin{figure}[t]
\centering
\includegraphics[width=1\linewidth]{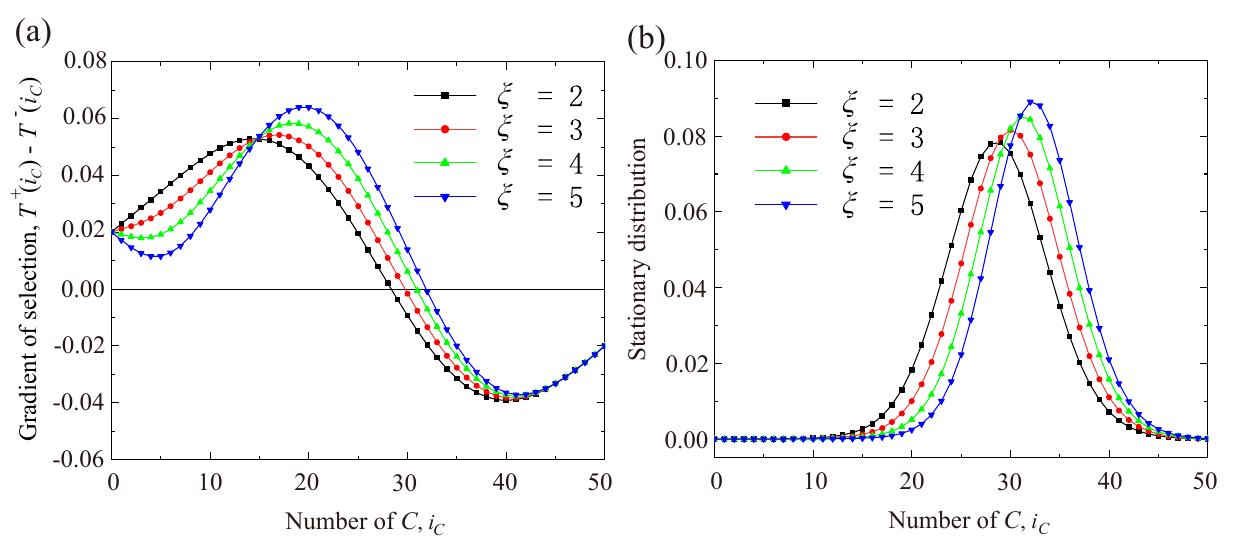}
\caption{Gradient of selection and stationary distributions for differen values of $\xi$ in finite populations. The presence of behavioral mutations makes it more advantageous to delay the adjustment of strategies for the emergence of high-level cooperation. Parameters are $Z=50, N=6, n_{pg}=3, T=0.2, r=0.3, w=0.8, \mu=1/Z, \beta=2$, and $\sigma=0.3$.}
\label{fig5}
\end{figure}

In Fig. 5, we present the impact of $\xi$ values on the evolutionary outcomes in finite populations. We find that for four different $\xi$ values, the gradient equation always exhibits a stable internal equilibrium point. Moreover, as the $\xi$ value increases, the equilibrium point moves towards the right axis, indicating higher levels of cooperation (see Fig. 5(a)). This deviation from the results in infinite populations is fundamentally attributed to stochastic effects. Results from the stationary distributions show that the population spends most of the time in configurations where cooperators prevail (see Fig. 5(b)). A higher value of $\xi$ ensures a higher level of cooperation.

When the population consists of cooperators, defectors, and conditional cooperators, we present in Fig. 6 the stationary distribution and gradient of selection for different values of $\xi$, when the risk is below the tolerance threshold and when the risk is above the tolerance threshold, respectively. We find that the population spends the most of time in configurations where defectors prevail when $r<T$. Additionally, an increase in $\xi$ does not alter the stochastic dynamics (see Fig. 6 (a)-(d)). This implies that defection is always favored by natural selection. When the risk is above the tolerance threshold, the stochastic dynamics of cooperation, defection, and conditional cooperation strategies will undergo significant changes. Specifically, when the value of $\xi$ is low, we find that the population spends a considerable amount of time in configurations in the center of the simplex, which means that the three strategies can stably coexist in the population (see Fig. 6 (e)). Most of the interior trajectories converge to this region, while a small number of trajectories converge to vertex $D$. As the value of $\xi$ increases, we observe that the shaded region moves towards the $D-C$ boundary, indicating a gradual decrease in the proportion of conditional cooperators in the steady state (see Fig. 6 (e)-(h)).

\begin{figure}[t]
\centering
\includegraphics[width=1\linewidth]{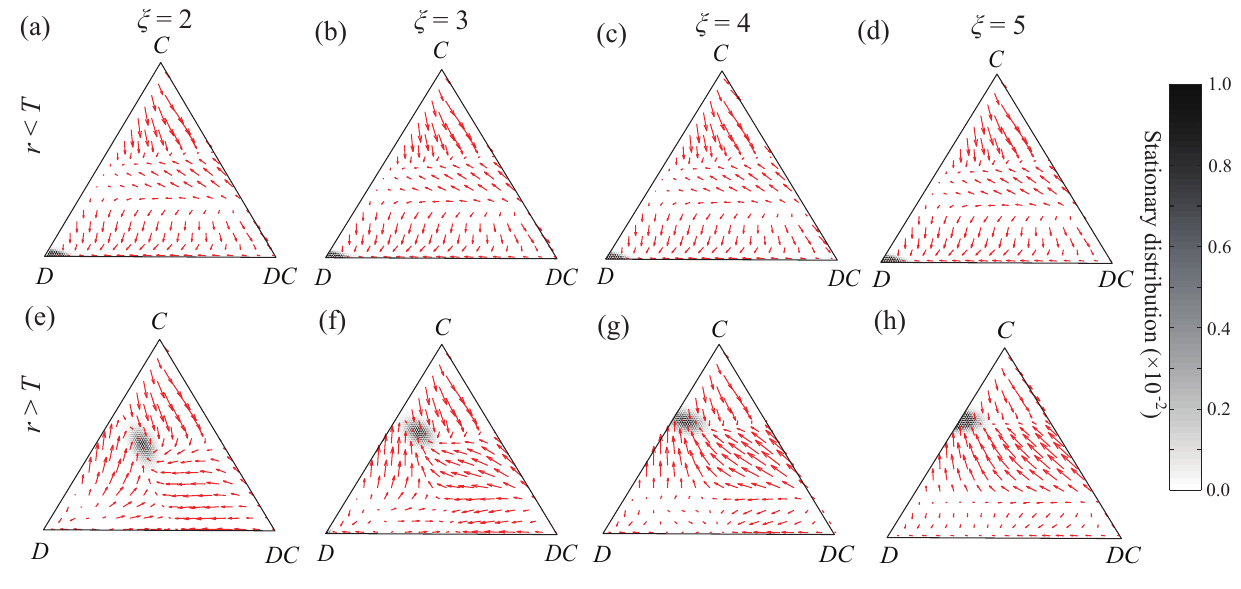}
\caption{The evolutionary dynamics of cooperation, defection, and conditional cooperation strategies in finite populations for different values of $\xi$. The simplex $S_{3}$ contains all possible configurations, where each small dot represents one state. The colorbar indicates the magnitude of the stationary distribution value and darker dots indicate those configurations visited more often. The red arrow is derived from the gradient equation solution and is used to describe the most likely evolutionary direction when leaving the current configuration. When the risk is below the tolerance threshold, defectors dominate the entire group, while when the risk exceeds the tolerance threshold, cooperators, defectors, and conditional cooperators can stably coexist in the population. Parameters are $Z=100, \mu=1/Z, \beta=5, N=6, n_{pg}=3, c=0.1, b=1, T=0.5, r=0.3, w=0.8, \sigma=0.1,$ and $\xi=2$ in panel (a); $\xi=3$ in panel (b); $\xi=4$ in panel (c); $\xi=5$ in panel (d); $Z=100, \mu=1/Z, \beta=5, N=6, n_{pg}=3, c=0.1, b=1, T=0.2, r=0.5, w=0.8, \sigma=0.1,$ and $\xi=2$ in panel (e); $\xi=3$ in panel (f); $\xi=4$ in panel (g); $\xi=5$ in panel (h).}
\label{fig6}
\end{figure}

We further investigate the impact of different model parameters on the stochastic dynamics of finite populations. In Fig. A5, we present outcomes from stationary distribution and gradient of selection for four different risk values. We find that when the risk is low, the population spends the majority of time in configurations where defectors prevail (see Fig. A5 (a)), and a slight increase in risk does not significantly alter the stationary distribution results but does affect the gradient of selection (two equilibrium states will appear on the $D-C$ boundary (see Fig. A5(b))). Continued increase in risk drives the emergence of interior steady states. As shown in Fig. A5(c), most of the interior trajectories flow towards the interior equilibrium states, a small portion flows towards vertex $D$, and the remaining trajectories converge to the middle portion of the $D-DC$ boundary. Such dynamical outcome persists until relatively large risk values (see Fig. A5(d)), which implies that high levels of risk can effectively enhance cooperation.

We are also interested in investigating the impact of different collective target values on the evolutionary dynamics when stochastic factors are taken into account. In the top row of Fig. A6, we study the effects of different collective targets on the evolutionary dynamics of cooperation, defection, and conditional cooperation when the risk is below the tolerance threshold. When the collective target value is small ($n_{pg}=2$), we observe there exist bistability, where the population spends time in both configurations where defectors win and configurations near the $D-C$ boundary (implying stable coexistence of cooperators and defectors within the population) (see Fig. A6(a)). As the collective target value increases, we observe that the population spends most of time in the configuration where the defectors prevail (see Fig. A6(b)-(d)). When the risk exceeds the tolerance threshold, we observe that for lower collective targets, the population spends more time inside the simplex where three strategies can coexist (see Fig. A6(e)). The increase in collective target values drives the shadow area to move towards configurations that favor cooperation (see Fig. A6(f)-(h)).

Furthermore, we are highly interested in conducting further research to examine the effects of observation costs on evolutionary outcomes in finite populations. We find that when the risk is below the tolerance threshold, for different observation cost values, the population spends most of the time in the configuration dominated by defectors (see Fig. A7(a)-(d)). However, when the risk is higher than the tolerance threshold, an interior coexistence state can emerge and gradually approach the $D-C$ boundary with the increase of observation costs (see Fig. A7(e)-(h)). This indicates that an increase in risk can facilitate the maintenance of high levels of cooperation. Finally, we aim to investigate the impact of the repeated probability of the game in finite populations on evolutionary outcomes (see Fig. A8). Similar to the results in infinite populations, we find that when the risk is below the tolerance threshold, defectors are favored by natural selection (see Fig. A8(a)-(d)); while when the risk is above the threshold, the system exhibits an interior steady state where cooperators, defectors, and conditional cooperators can coexist (see Fig. A8(e)-(h)). Therefore, the increase in the repeated probability does not significantly alter the dynamics of the system.

\vbox{}
\leftline{\textbf{Conclusions}}

In collective-risk social dilemma games, conflicts of interest among individuals often hinder cooperation, leading to a loss of overall benefits. To address this issue, we explored the collective-risk social dilemma through a theoretical model incorporating repeated group interactions and a conditional cooperation strategy. This strategy allows defectors to assess the game's risk level and the number of cooperators before deciding whether to contribute in later rounds. Our key findings reveal that introducing a conditional cooperation strategy can foster cooperation, even in situations of relatively low collective risk.

Conditional cooperation strategies are widely observed in the framework of evolutionary game theory. Such strategies allow individuals to adjust their actions based on the behavior of others in the group and the game environment they are in. For example, in the context of emissions reduction actions, major carbon-emitting countries including the United States, the European Union, Japan, China, and India, have stated in a series of declarations that they will only significantly reduce emissions if other countries take corresponding actions \cite{Tingley14}. Besides, previous experimental and theoretical studies based on collective-risk social dilemma games have demonstrated that individuals tend to contribute based on the previous decisions made by other group members \cite{Domingos20,Greenwood11}. The findings indicate that conditional behavior, also called trigger strategies in \cite{Greenwood11}, is more successful than those adopting fixed strategies. Previous studies on collective risk social dilemma games have considered conditional strategies (individuals make decisions based on the total contributions) and different perceived risk factors \cite{Abou2018,JTheoreticalBiology_2014}. The results indicate that by adding loss parameters, one can shift the system from late to early contributors. Inspired by these studies, we introduce $DC$ strategists who only switch in the last rounds, but it isn't clear if it will happen, which adds another level of uncertainty compared to previous studies.

The emergence of conditional cooperation strategies depends on the repeated group interactions scenario. In repeated games, players have the opportunity to observe each other's behavior in multiple interactions and adjust their strategies based on their observations. Previous theoretical studies on collective-risk social dilemmas primarily relies on one-shot game interaction scenarios where participants are only able to make a single decision \cite{Santos2012,Chen2012,chen2014}. As a result, participants often opt for selfish strategies, leading to the breakdown of cooperation and the emergence of social dilemmas. In this work, we consider the repeated groups interactions where individuals can adjust their behavior based on the game environment, including the risk level and group cooperation level. Our findings demonstrate that the risk level and the number of game rounds for strategy adjustment have a significant impact on the evolutionary outcomes. Specifically, when the risk level exceeds the tolerance threshold of conditional cooperators, timely strategy adjustment by conditional cooperators can promote the stable coexistence of cooperators, defectors, and conditional cooperators in the population. Moreover, an increase in the number of round for strategy adjustment expands the prevalence of free-riders in the population.

In our work, we assume that conditional cooperators decide whether to cooperate based on their assessment of the cooperation level within the group and game risk level. It should be noted that both the cooperation level and risk level are evaluated by the conditional cooperators based on the initial rounds of the game, and there may be errors in the evaluation \cite{Santos21}. Therefore, it is worthwhile to investigate the impact of overestimation and underestimation on the evolutionary outcome. Furthermore, decision-making by individuals often involves decision cost, and it is worth further exploring how the introduction of decision cost affects the evolutionary outcomes of the system. Risk can affect cooperative behavior in various ways \cite{Liu23elife,Hagel16sr}, and in our model, the risk value of the game is constant. However, in reality, the shape of the risk curve may be diverse. Abou Chakra and coworkers \cite{Abou2018} manipulated the shape of the risk curve and investigated how risk curve characteristics affect individual contributions. Therefore, future research could investigate the impact of different risk equations on the evolutionary outcomes. Finally, it is important to emphasize that the study of cooperation is just one type of moral behavior that can be studied using these models \cite{Capraro21interface}. To better understand selfless human behavior and adjust policies accordingly, future research could focus on mathematical modeling of moral preferences. This represents a promising avenue for advancing our understanding of moral decision-making and promoting socially responsible behavior.

Individual decision-making, oscillating between cooperative and defective strategies, is often governed by the balance of risk against personal tolerance thresholds and the achievement of collective objectives. This is a ubiquitous phenomenon observed across various domains. Taking climate change as an illustrative example, if the collective objective of mitigating climate change has not yet been achieved and the perceived risks surpasses a nation's tolerance level, the country is more likely to adopt cooperative strategies. These may include reducing greenhouse gas emissions, investing in renewable energy, and supporting international climate agreements. By constructing a theoretical model, our work reveals that timely transition from defective strategy to cooperative strategy is beneficial for the construction of a cooperative society.

\clearpage

\noindent \textbf{Acknowledgments} \\
This research was supported by the Natural Science Foundation of Shaanxi (Grant No. 2023-JC-QN-0791) and the Fundamental Research Funds of the Central Universities of China (Grants Nos. 2452022012, 2452022144).\\

\noindent \textbf{Data accessibility}\\
Source code is available at the Dryad, Dataset, https://doi.org/10.5061/dryad.80gb5mkw0.

\noindent \\ \textbf{Author contributions} \\
S. H.: formal analysis, investigation, methodology, writing—original draft, writing—review and editing; 
Z. H.: formal analysis, writing—original draft;
L. L.: conceptualization, investigation, formal analysis, funding acquisition, supervision, writing—original draft, writing—review and editing.

\noindent \\ \textbf{Competing financial interests} \\
The authors declare no competing financial interests.


\begin{thebibliography}{54}

\bibitem{Perc_PR_17}
M. Perc, J. J. Jordan, D. G. Rand, Z. Wang, S. Boccaletti, A. Szolnoki, Statistical physics of human cooperation. Physics Reports, 687(2017) 1-51.

\bibitem{Panis_22}
J. Lie-Panis, J. B. André, Cooperation as a signal of time preferences. Proceedings of the Royal Society B: Biological Sciences, 289(2022), 20212266.

\bibitem{Stockley_11}
P. Stockley, J. Bro‐Jørgensen, Female competition and its evolutionary consequences in mammals. Biological Reviews, 86(2011), 341-366.
\bibitem{Thomas_04}
C. D. Thomas, A. Cameron, R. E. Green, et al., Extinction risk from climate change. Nature, 427(2004), 145-148.


\bibitem{Boesch_94}
C. Boesch, Cooperative hunting in wild chimpanzees. Animal Behaviour, 48(1994), 653-667.

\bibitem{Ranta_93}
E. Ranta, H. Rita, K. Lindstrom, Competition versus cooperation: success of individuals foraging alone and in groups. The American Naturalist, 142(1993), 42-58.

\bibitem{Wilkinson_16}
G. S. Wilkinson, G. G. Carter, K. M. Bohn, D. M. Adams, Non-kin cooperation in bats. Philosophical Transactions of the Royal Society B: Biological Sciences, 371(2016), 20150095.



\bibitem{chen_22prsa}
X. Chen, F. Fu, Highly coordinated nationwide massive travel restrictions are central to effective mitigation and control of COVID-19 outbreaks in China. Proceedings of the Royal Society A: Mathematical Physical and Engineering Sciences, 478(2022), 20220040.

\bibitem{chen_19prsb}
X. Chen, F. Fu, Imperfect vaccine and hysteresis. Proceedings of the Royal Society B: Biological Sciences, 286(2019), 20182406.

\bibitem{Andrews_18}
T. M. Andrews, A. W. Delton, R. Kline, High-risk high-reward investments to mitigate climate change. Nature Climate Change, 8(2018), 890-894.

\bibitem{Grimalda_2022}
G. Grimalda, A. Belianin, H. Hennig-Schmidt, T. Requate, M. V. Ryzhkova, Sanctions and international interaction improve cooperation to avert climate change. Proceedings of the Royal Society B: Biological Sciences, 289(2022), 20212174.

\bibitem{Schlter_16}
M. Schlüter, A. Tavoni, S. Levin, Robustness of norm-driven cooperation in the commons. Proceedings of the Royal Society B: Biological Sciences, 283(2016), 20152431.

\bibitem{Bornstein_94}
G. Bornstein, M. Ben-Yossef, Cooperation in intergroup and single-group social dilemmas. Journal of Experimental Social Psychology, 30(1994), 52-67.

\bibitem{Liu_22interface}
L. Liu, Z. Xiao, X. Chen, A. Szolnoki, Early exclusion leads to cyclical cooperation in repeated group interactions. Journal of the Royal Society Interface, 19(2022), 20210755.


\bibitem{han2015}
T. A. Han, C. Perret, S. T. Powers, When to (or not to) trust intelligent machines: Insights from an evolutionary game theory analysis of trust in repeated games. Cognitive Systems Research, 68(2021) 111-124.

\bibitem{han2021}
T. A. Han, L. M. Pereira, T. Lenaerts, F. C. Santos, Mediating artificial intelligence developments through negative and positive incentives. PLoS ONE, 16(2021) e0244592.

\bibitem{han2022}
T. A. Han, Emergent behaviours in multi-agent systems with Evolutionary Game Theory. AI Communications, 35(2022) 327-337.

\bibitem{han2022interface}
T. A. Han, Institutional incentives for the evolution of committed cooperation: ensuring participation is as important as enhancing compliance. Journal of the Royal Society Interface, 19(2022) 20220036.


\bibitem{Tanimoto2017}
J. Tanimoto, How does resolution of strategy affect network reciprocity in spatial prisoner's dilemma games? Applied Mathematics and Computation, 301(2017), 36-42.

\bibitem{Tanimoto2015}
J. Tanimoto, \emph{Fundamentals of Evolutionary Game Theory and its Applications}. Springer Japan, (2015).

\bibitem{Tanimoto2021}
J. Tanimoto, \emph{Sociophysics Approach to Epidemics}. Singapore: Springer, (2021).

\bibitem{Szolnoki2014interface}
A. Szolnoki, M. Mobilia, L. L. Jiang, B. Szczesny, A. M. Rucklidge, M. Perc, Cyclic dominance in evolutionary games: a review. Journal of the Royal Society Interface, 11(2014) 20140735.

\bibitem{Szolnoki2008epl}
A. Szolnoki, M. Perc, Z. Danku, Making new connections towards cooperation in the prisoner's dilemma game. EPL, 84(2008), 50007.

\bibitem{Xia2018NJP}
C. Xia, X. Li, Z. Wang, M. Perc, Doubly effects of information sharing on interdependent network reciprocity. New Journal of Physics, 20(2018), 075005.

\bibitem{Zhu2022}
Y. Zhu, C. Xia, Z. Wang, Z. Chen, Networked decision-making dynamics based on fair, extortionate and generous strategies in iterated public goods games. IEEE Transactions on Network Science and Engineering, 9(2022), 2450-2462.

\bibitem{xia2022}
C. Xia, Z. Hu, D. Zhao, Costly reputation building still promotes the collective trust within the networked population. New Journal of Physics, 24(2022), 083041.

\bibitem{Chen2012}
X. Chen, A. Szolnoki, M. Perc, Risk-driven migration and the collective-risk social dilemma. Physical Review E, 86(2012), 036101.

\bibitem{He2019}
N. He, X. Chen, A. Szolnoki, Central governance based on monitoring and reporting solves the collective-risk social dilemma. Applied Mathematics and Computation, 347(2019), 334-341.

\bibitem{chen2014}
X. Chen, Y. Zhang, T. Z. Huang, M. Perc, Solving the collective-risk social dilemma with risky assets in well-mixed and structured populations. Physical Review E, 90(2014), 052823.

\bibitem{chen2012epl}
X. Chen, A. Szolnoki, M. Perc, Averting group failures in collective-risk social dilemmas. EPL, 99(2012), 68003.

\bibitem{wang2009PRE}
J. Wang, F. Fu, T. Wu, L. Wang, Emergence of social cooperation in threshold public goods games with collective risk. Physical Review E, 80(2009), 016101.

\bibitem{Milinski2008}
M. Milinski, R. D. Sommerfeld, H. J. Krambeck, F. A. Reed, J. Marotzke, The collective-risk social dilemma and the prevention of simulated dangerous climate change. Proceedings of the National Academy of Sciences, 105(2008), 2291-2294.

\bibitem{Santos2012}
F. C. Santos, V. V. Vasconcelos, M. D. Santos, P. N. B. Neves, J. M. Pacheco, Evolutionary dynamics of climate change under collective-risk dilemmas. Mathematical Models and Methods in Applied Sciences, 22(2012), 1140004.

\bibitem{Szekely2021}
A. Szekely, F. Lipari, A. Antonioni, M. Paolucci, A. Sánchez, L. Tummolini, G. Andrighetto, Evidence from a long-term experiment that collective risks change social norms and promote cooperation. Nature Communications, 12(2021), 5452.

\bibitem{Abou2018}
M. Abou Chakra, S. Bumann, H. Schenk, A. Oschlies, A. Traulsen, Immediate action is the best strategy when facing uncertain climate change. Nature Communications, 9(2018), 2566.

\bibitem{santos_11}
F. C. Santos, J. M. Pacheco, Risk of collective failure provides an escape from the tragedy of the commons. Proceedings of the National Academy of Sciences, 108(2011) 10421-10425.

\bibitem{Liu2018}
L. Liu, X. Chen, Evolution of public cooperation in a risky society with heterogeneous assets. Frontiers in Physics, 5(2018), 67.

\bibitem{sun2021iScience}
W. Sun, L. Liu, X. Chen, A. Szolnoki, V. V. Vasconcelos, Combination of institutional incentives for cooperative governance of risky commons. iScience, 24(2021), 102844.

\bibitem{gis2019sr}
A. R. Góis, F. P. Santos, J. M. Pacheco, F. C. Santos, Reward and punishment in climate change dilemmas. Scientific Reports, 9(2019), 16193.

\bibitem{Vasconcelos2014pnas}
V. V. Vasconcelos, F. C. Santos, J. M. Pacheco, S. A. Levin, Climate policies under wealth inequality. Proceedings of the National Academy of Sciences, 111(2014), 2212-2216.
\bibitem{Jiang2023chaos}
L. L. Jiang, Z. Chen, M. Perc, Z. Wang, J. Kurths, Y. Moreno, Deterrence through punishment can resolve collective risk dilemmas in carbon emission games. Chaos: An Interdisciplinary Journal of Nonlinear Science, 33(2023), 043127.
\bibitem{Wang2020pnas}
Z. Wang, M. Jusup, H. Guo, L. Shi, S. Geček, M. Anand, M. Perc, C. T. Bauch, J. Kurths, S. Boccaletti, H. J. Schellnhuber, Communicating sentiment and outlook reverses inaction against collective risks. Proceedings of the National Academy of Sciences, 117(2020), 17650-17655.

\bibitem{Van Segbroeck_12PRL}
S. Van Segbroeck, J. M. Pacheco, T. Lenaerts, F. C. Santos, Emergence of fairness in repeated group interactions. Physical Review Letters, 108(2012), 158104.




\bibitem{Liu_prsa_22}
L. Liu, X. Chen, Indirect exclusion can promote cooperation in repeated group interactions. Proceedings of the Royal Society A: Mathematical Physical and Engineering Sciences, 478(2022), 20220290.
\bibitem{Hilbe13plos}
C. Hilbe, M. Abou Chakra, P. M. Altrock, A. Traulsen, The evolution of strategic timing in collective-risk dilemmas. PLoS ONE, 8(2013), e66490.
\bibitem{Chakra12pcb}
M. Abou Chakra, A. Traulsen, Evolutionary dynamics of strategic behavior in a collective-risk dilemma. PLoS Computational Biology, 8(2012), e1002652.

\bibitem{Tingley_14}
D. Tingley, M. Tomz, Conditional cooperation and climate change. Comparative Political Studies, 47(2014), 344-368.
\bibitem{JTheoreticalBiology_2014}
M. Abou Chakra, A. Traulsen, Under high stakes and uncertainty the rich should lend the poor a helping hand. Journal of Theoretical Biology, 341(2014) 123-130.


\bibitem{Sigmund10CS}
K. Sigmund, \emph{The Calculus of Selfishness}. Princeton University Press, Princeton, NJ, (2010).


\bibitem{Schuster_83}
P. Schuster, K. Sigmund, Replicator dynamics. Journal of Theoretical Biology, 100(1983) 533-538.

\bibitem{Hofbauer_98}
J. Hofbauer, K. Sigmund, \emph{Evolutionary Games and Population Dynamics}. Cambridge University Press, Cambridge, UK, (1998).


\bibitem{Imhof_05}
L. A. Imhof, D. Fudenberg, M. A. Nowak, Evolutionary cycles of cooperation and defection. Proceedings of the National Academy of Sciences, 102(2005), 10797-10800.

\bibitem{Kampen_07}
N. G. Van Kampen, \emph{Stochastic Processes in Physics and Chemistry}. NorthHolland: NorthHolland Personal Library, (2007).

\bibitem{Vasconcelos_13}
V. V. Vasconcelos, F. C. Santos, J. M. Pacheco, A bottom-up institutional approach to cooperative governance of risky commons. Nature Climate Change, 3(2013), 797-801.

\bibitem{szab_98}
G. Szabó, C. Tőke, Evolutionary prisoner's dilemma game on a square lattice. Physical Review E, 58(1998), 69.


\bibitem{Tingley14}
D. Tingley, M. Tomz, Conditional cooperation and climate change. Comparative Political Studies, 47(2014), 344-368.

\bibitem{Domingos20}
E. F. Domingos, J. Grujić, J. C. Burguillo, G. Kirchsteiger, F. C. Santos, T. Lenaerts, Timing uncertainty in collective risk dilemmas encourages group reciprocation and polarization. iScience, 23(2020), 101752.
\bibitem{Greenwood11}
G. Greenwood, Evolution of strategies for the collective-risk social dilemma relating to climate change. EPL, 95(2011), 40006.

\bibitem{Santos21}
F. P. Santos, S. A. Levin, V. V. Vasconcelos, Biased perceptions explain collective action deadlocks and suggest new mechanisms to prompt cooperation. iScience, 24(2021), 102375.

\bibitem{Liu23elife}
L. Liu, X. Chen, A. Szolnoki, Coevolutionary dynamics via adaptive feedback in collective-risk social dilemma game. eLife, 12(2023), e82954.
\bibitem{Hagel16sr}
K. Hagel, M. Abou Chakra, B. Bauer, A. Traulsen, Which risk scenarios can drive the emergence of costly cooperation?. Scientific Reports, 6(2016), 19269.

\bibitem{Capraro21interface}
V. Capraro, M. Perc, Mathematical foundations of moral preferences. Journal of the Royal Society Interface, 18(2021), 20200880.


\end{thebibliography}
\end{document}